\documentstyle[aps,eqsecnum,preprint,tighten,floats]{revtex}

\def\be{\begin{equation}}
\def\ee{\end{equation}}

\begin{document}
\draft


\preprint{\vbox{\it 
                        \null\hfill\rm    SI-TH-96-4, hep-ph/9607367}\\\\}
%
\title{Electro-excitation amplitudes of the $\Delta$ - isobar\\
in the Skyrme model}
\author{ H.~Walliser and G.~Holzwarth\thanks{%
e-mail: holzwarth@hrz.uni-siegen.d400.de}}
\address{Fachbereich Physik, Universit\"{a}t-GH-Siegen, 
D-57068 Siegen, Germany} 
%
%
\maketitle
\begin{abstract}
Electro magnetic transition form factors for the excitation of the 
$\Delta 33$-resonance are evaluated in the Skyrme model. They
crucially rely on rotationally induced deformations of the hedgehog 
soliton which are suppressed by two $N_C$-orders as compared to the
leading parts of the isovector current. Partial photon coupling through
vector mesons is included in a schematic way. Recoil corrections
are approximated by a boost to the equal-velocity frame. The results
for the photodecay amplitudes agree with  
experimental numbers and the shapes of $M1, E2, C2 -$ transition form
factors show essential features as observed in electro-excitation
experiments.   
\end{abstract}
%

\newpage


\section{Introduction}

Electro magnetic transition form factors for the excitation of nucleon
resonances present challenging terrain for nucleon models because they
sensitively reflect the nature of the states excited by the virtual
photon. For the most prominent nucleon resonance, the $\Delta
(1232)$, although existing data are sparse, there is sufficient
indication that the magnetic ($M1$) transition form factor
$G_{M1}^{N\Delta}$ is
significantly different from the elastic proton magnetic form factor $G_M^P$.
Up to the highest measured values of momentum transfer $Q^2$ 
both form factors decrease relative to
the standard dipole shape $G_D = (1+Q^2/0.71 GeV^2)^{-2}$. However, the
decrease of the transition form factor sets in much earlier and with larger
slope such that near 5 $GeV^2$ the ratio $G_{M1}^{N\Delta} /G_D$ has
dropped to about half of its value at $Q^2$=0 while $G_M^P/ \mu^P G_D$ is
still close to one. On the theoretical side a recent extensive analysis
in terms of a relativized quark model \cite{wspr90} not only has
problems with obtaining a correct value of the $M1$ transition moment 
at $Q^2=0$ but shows also a severe dependence of the shape of 
$G_{M1}^{N\Delta}$ on the quark wave functions and configuration mixing.

For the transverse electric ($E2$) and longitudinal ($C2$) form factors
the experimental information is even more rudimentary. From the point
of perturbative QCD one would expect asymptotic equality of  
$E2$ and $M1$ transition amplitudes
\cite{pb80} which would imply a change
of sign in $G_{E2}^{N\Delta}$ for $Q^2$ higher than the values where it is
presently known to be much smaller and of opposite sign relative to
$G_{M1}^{N\Delta}$. There are experimental indications that this sign
change occurs near $1.5 - 2~ GeV$ \cite{dmw91}. 

Solitons in effective nonlinear meson field theories present an
attractive alternative for the evaluation of baryon form factors,
because already in the leading classical approximation of ${\cal
O}(N_C)$ ($N_C$ is the number of colours)
the spatial structure of currents is determined through 
the classical solution for the soliton profile which in the language of
chiral perturbation theory sums up all multi-loop 
graphs without closed meson loops \cite{m94} \cite{dm94}. 
Transition moments and form factors in leading order $N_C$ have been
very early discussed in \cite{anw83} \cite{an85} \cite{ww87}.
However, it is evident that for these observables the 
${\cal O}(N_C)$-approximation is not sufficient: i) In classical
approximation nucleon and $\Delta$ are characterized by the same
soliton profile therefore transition matrix elements can differ from
diagonal matrix elements only by geometrical factors, i.e. normalized
form factors coincide; ii) due to the spherical symmetry of the classical
hedgehog soliton quadrupole matrix elements are zero; this implies 
vanishing $E2$ and $C2$ form factors; iii) longitudinal matrix elements
are related to the time-component of the vector current; in the
equation of continuity for the vector current the contribution of the 
time component is suppressed by $1/N_C^2$ as compared to the leading
part of the spatial components; 
current conservation therefore requires solving the equations of motion 
consistently to ${\cal O}(N_C^{-1})$. These rotational contributions
to the isovector form factors are of the same ${\cal O}(N_C^{-1})$ as
the isoscalar part of the magnetic form factor which always had been
included in the evaluation of nucleon magnetic properties.

It has been shown \cite{mw96} that inclusion of ${\cal
O}(N_C^{-1})$-rotational effects in the equations of motion 
introduces quadrupole distortion into the soliton. We show in section
III that the resulting structure of the currents in terms of collective
operators produces nonvanishing ${\cal O}(N_C^{-1})$
$E2$ and $C2$ form factors and a nonvanishing difference between 
elastic and transition magnetic form factors. To demonstrate the 
essential features of their shape we evaluate them for the most
simple soliton, the skyrmion. For a realistic description of the
photon-baryon coupling the Skyrme model should be augmented by vector
mesons. We include their influence in an approximate form by one
common vector meson propagator. 

In the most interesting region of $Q^2$-values above 1 $GeV^2$ the
shape of the form factors is sensitive to relativistic recoil
corrections. Their reliable inclusion poses a serious problem 
in quark models as well as in soliton models. Following \cite{j91}
\cite{fh82}
we perform a boost of the soliton to the equal-velocity-frame.
Naturally, its effect depends heavily on the kinematical mass of the
soliton which (in tree approximation) exceeds the actual 
nucleon mass by up to a factor of two. Therefore the resulting 
shape of the form factors for higher $Q^2$ is not really reliable
but only indicative of the expected behaviour. It seems that also in
this respect the explicit inclusion of vector mesons appears very
helpful, because it leads to a sizable lowering of the soliton mass.

In section IV we present the results of this rather simple model
of nucleon and $\Delta$-resonance for the helicity amplitudes at the
photon point and the transition form factors.

\section{Definitions}

In this section we establish the connection between the helicity
amplitudes and the $M1, E2$ and $C2$ multipole operators which may
contribute to the transition from the proton to the $\Delta$-isobar.
We also define the form factor conventions used in this paper.

In order to minimize errors due to recoil, 
the so-called "equal-velocity" (EV) frame where
the incoming nucleon and the outgoing $\Delta$ have opposite velocities
\be
\label{veloce}
v^2_\Delta = v^2_N = v^2
= \frac{q^2}{q^2+(M_\Delta+M_N)^2} \; , \quad
\gamma^2 = \frac{1}{1-v^2}
\ee
is chosen as a convenient reference frame \cite{wspr90}. Here
$\mbox{\boldmath$q$}$ represents the three-momentum of the virtual
photon in the EV-frame
\begin{eqnarray}
\label{phomom}
&& q^2 = \mbox{\boldmath$q$}^2 = \frac{(M_\Delta+M_N)^2}{4M_\Delta M_N}
\left[ (M_\Delta-M_N)^2 + Q^2 \right] 
\nonumber \\
&& q^2_0 = \mbox{\boldmath$q$}^2 - Q^2 \; .
\end{eqnarray}
For elastic scattering ($M_\Delta=M_N$) this frame
reduces to the Breit frame $\mbox{\boldmath$q$}^2 = Q^2$, $q_0=0$
and also for $Q$ large compared to the nucleon-$\Delta$ split
we have $\mbox{\boldmath$q$}^2 \simeq Q^2$, whereas at
the photon point ($Q^2=0$) we obtain 
$q_0=|\mbox{\boldmath$q$}|=q_\Delta = (M_\Delta^2 - M_N^2)/2
\sqrt{M_\Delta M_N} = 296 MeV$.

Concerning the helicity amplitudes, it is convenient to decompose the
transverse ones
\begin{eqnarray}
\label{helix}
&& A_{\frac{1}{2}} = A_{\frac{1}{2}} (M1) + A_{\frac{1}{2}} (E2) 
= - \frac{1}{2} (M1 + 3E2) \\ 
&& A_{\frac{3}{2}} = A_{\frac{3}{2}} (M1) + A_{\frac{3}{2}} (E2)
= \sqrt{3} A_{\frac{1}{2}} (M1) 
- \frac{1}{\sqrt{3}} A_{\frac{1}{2}} (E2) 
= - \frac{\sqrt{3}}{2} (M1 - E2)
\nonumber
\end{eqnarray}
into their $M1$ and $E2$ contributions. Then all helicity amplitudes
may be expressed by simple matrix elements 
\begin{eqnarray}
\label{helix2}
&&A_{\frac{1}{2}} (M1) = \sqrt{\frac{4 \pi \alpha}{2 k_\Delta}}
<\Delta ^{+},S_3=\frac{1}{2}|M^{M1}_{\lambda=1}|p,S_3=-\frac{1}{2}>
\nonumber \\
&&A_{\frac{1}{2}} (E2) = \sqrt{\frac{4 \pi \alpha}{2 k_\Delta}}
<\Delta ^{+},S_3=\frac{1}{2}|M^{E2}_{\lambda=1}|p,S_3=-\frac{1}{2}> \\
&&S_{\frac{1}{2}} (C2) = \sqrt{\frac{4 \pi \alpha}{2 k_\Delta}}
\frac{M_N}{M_\Delta} \frac{q_0}{k_\Delta}
<\Delta ^{+},S_3=\frac{1}{2}|M^{C2}_{\lambda=0}|p,S_3=\frac{1}{2}>
\nonumber
\end{eqnarray}
($\alpha=1/137, k_\Delta=(M_\Delta^2-M_N^2)/2M_\Delta$) of the
corresponding multipole operators \cite{fw66}
\begin{eqnarray}
\label{mupo}
&&M^{M1}_\lambda (q^2) = i \sqrt{6 \pi} \lambda \int d^3\!r\, V^3_i \, j_1(qr) \left[
\mbox{\boldmath$Y$}_{11\lambda}( \mbox{\boldmath$ \hat{r}$} ) \right]_i
\nonumber \\
&&M^{E2}_\lambda (q^2) = \frac{\sqrt{10 \pi}}{q} \int d^3\!r\, V^3_i \, \left[
\mbox{\boldmath$\nabla$} \times 
(j_2(qr) \mbox{\boldmath$Y$}_{22\lambda}( \mbox{\boldmath$
\hat{r}$} )) \right]_i \\ 
&&M^{C2}_\lambda (q^2) = -\sqrt{20 \pi} \int d^3\!r\, V^3_0 \, 
j_2(qr) Y_{2 \lambda}( \mbox{\boldmath$\hat{r}$} ) \; .
\nonumber
\end{eqnarray}
Note here, that for the nucleon-$\Delta$ transition only
the isovector piece $V^a_\mu$ of the electromagnetic
current can contribute.  
Finally we introduce the transition form factors 
\begin{eqnarray}
\label{fofa}
&&<\Delta ^{+},S_3=\frac{1}{2}|M^{M1}_{\lambda=1}(q^2)|p,S_3=-\frac{1}{2}>
= - \frac{q}{2 \sqrt{2} M_N} G^{N \Delta}_{M1}(q^2)  \nonumber \\
&&<\Delta ^{+},S_3=\frac{1}{2}|M^{E2}_{\lambda=1}(q^2)|p,S_3=-\frac{1}{2}>
= - \frac{3q_0 q}{4 \sqrt{2}} G^{N \Delta}_{E2}(q^2) \\
&&<\Delta ^{+},S_3=\frac{1}{2}|M^{C2}_{\lambda=0}(q^2)|p,S_3=\frac{1}{2}>
= - \frac{q^2}{2} G^{N \Delta}_{C2}(q^2) \; . \nonumber
\end{eqnarray}
The normalization is chosen such that the form factors
at $q^2=0$ are equal to the corresponding transition
magnetic and quadrupole moments. The helicity amplitudes
(\ref{helix},\ref{helix2}) are readily expressed by these form factors
as well as the electromagnetic ratio and the ratio between the
longitudinal and transverse couplings
\be
\label{ratio}
\frac{E2}{M1} 
= \frac{1}{3} \frac{A_{\frac{1}{2}}(E2)}{A_{\frac{1}{2}}(M1)} 
= \frac{A_{\frac{1}{2}}
-\frac{1}{\sqrt{3}} A_{\frac{3}{2}}}
{A_{\frac{1}{2}} + \sqrt{3} A_{\frac{3}{2}}} \; , \qquad 
\frac{C2}{M1} 
= -\frac{1}{\sqrt{2}} \frac{S_{\frac{1}{2}}(C2)}{M1} 
= \frac{\sqrt{2} S_{\frac{1}{2}}}
{A_{\frac{1}{2}} + \sqrt{3} A_{\frac{3}{2}}} \;
\ee
(our longitudinal amplitude $S_{\frac{1}{2}}$ in eq.(\ref{helix2})
differs by a factor $-1/\sqrt{2}$ from the one used in \cite{bm87} \cite{ra88}).
The definitions of the multipole operators (\ref{mupo}) and form factors
(\ref{fofa}) involve the components $V_\mu^a$ of the vector current in
the EV - frame where the soliton is moving with velocity $v$ from 
eq.(\ref{veloce}). If we denote the form factors evaluated in the soliton
rest frame by $\tilde G$ the relativistically corrected form factors $G$
in the EV - frame are approximately obtained through the relations
\begin{eqnarray}
\label{relativ}
&& qG^{N \Delta}_{M1}(q^2) =  \frac{q}{\gamma^2}
\tilde G^{N \Delta}_{M1}(\frac{q^2}{\gamma^2}) 
\; , \nonumber \\
&& q_0 q G^{N \Delta}_{E2}(q^2)  =  \frac{q_0 q}{\gamma^2}
\tilde G^{N \Delta}_{E2}(\frac{q^2}{\gamma^2}) 
\; , \\
&& q^2 G^{N \Delta}_{C2}(q^2)  =  \frac{q^2}{\gamma^2}
\tilde G^{N \Delta}_{C2}(\frac{q^2}{\gamma^2}) 
\; . \nonumber 
\end{eqnarray}
It should be appreciated that this procedure for solitons has a
profound basis in the Lorentz-covariance of the underlying field theory:
The boosted soliton again is solution of the boosted
equations-of-motion.  This is in contrast to 'prescriptions'
in quark cluster model approaches which try to incorporate some relativity.
A deficiency lies in the fact that in the Lorentz-boost the collective
position and momentum variables are treated as commuting classical 
variables~\cite{j91}. In tree approximation, however, the whole soliton 
approach is basically classical, so on this level the boost 
appears consistent. The real problem lies, however, in the 
fact that quantum corrections apparently are large, 
especially for the soliton mass,
which through the Lorentz factors enters sensitively into
the behaviour of formfactors for large $q^2$.
But at tree level we have to accept this deficiency, therefore
in this respect the results can
only indicate general features and have no strong predictive power.

\section{Multipoles in the soliton model}

In this section we evaluate the $M1, E2$ and $C2$ multipole operators
in the most simple pseudoscalar soliton model, explicit expressions for the
corresponding transition form factors (\ref{fofa}) will 
subsequently be given. 

If we insert the hedgehog ansatz 
\be
\label{hedgehog}
U = A U_0 A^\dagger \; ,  \qquad  U_0 = e^{i \mbox{\boldmath$\tau \hat{r}$}
F(r)}
\ee
(rotation matrix $A \in SU(2)$, chiral angle $F(r)$)
into the spatial components of the
vector current we find a nonvanishing $M1$ contibution but
the $E2$ contribution vanishes identically. This is because the $E2$
transition is related to a quadrupole deformation which is
suppressed by $1/N_C^2$ as compared to the $M1$ transition and which is
not present in the hedgehog ansatz. This
deformation, caused by the soliton's rotation, has to be taken into account
on the same footing as for the $C2$ transition which is related to the
time component of the vector current and therefore must contain 
an angular velocity.
For that reason we have to solve the equation of motion 
$\partial^\mu V_\mu^a = 0$ consistently to order $1/N_C^2$. This will
be done in the following subsection.

\subsection{Rotationally induced soliton deformations}

Small (time independent) soliton deformations $\mbox{\boldmath$\eta$}$
are introduced via the ansatz
\be
\label{callan}
U = A \sqrt{U_0} e^{i \mbox{\boldmath$\tau \eta$}/f_\pi}
\sqrt{U_0} A^{\dagger} \; , \qquad
\mbox{\boldmath$\eta$} = \mbox{\boldmath$\hat{r}$} \eta_L
+ \mbox{\boldmath$\eta$}_T \; .
\ee
The driving term for these deformations (linear in
$\mbox{\boldmath$\eta$}$) stems from the centrifugal term in the
lagrangian which is proportional to the angular velocity
$\mbox{\boldmath$ \Omega$}^R$ squared 
\begin{eqnarray}
L_{\Omega}=f_{\pi} \int d^3\!r\,\, && \left\{ \left[ sc 
a_L - c_4^a \frac{1}{r^2}
(r^2 F^{\prime} s^2)^{\prime} \right] 
(\mbox{\boldmath$ \hat{r} \times
\Omega$}^R)^2 \eta_L   -s a_T
(\mbox{\boldmath$\hat{r}\Omega$}^R)
(\mbox{\boldmath$ \Omega$}^R \mbox{\boldmath$ \eta$}_T) 
\; \right\} \; .
\label{soldef}
\end{eqnarray}
Here and in the following we use the abbreviations $c_4^a=1/f^2_\pi e^2$ and 
\begin{eqnarray}
\label{metric}
& \displaystyle{a_L = 1+c^a_4 (F^{\prime 2} +  \frac{2s^2}{r^2})} \; , \quad
& \displaystyle{b_L = 1+c^a_4 \frac{2s^2}{r^2}}  \; , \nonumber \\
& \displaystyle{a_T = 1+c^a_4 (F^{\prime 2} -  \frac{2s^2}{r^2})} \; , \quad
& \displaystyle{b_T = 1+c^a_4 (F^{\prime 2} +  \frac{s^2}{r^2})} \; , \\
& \displaystyle{a_0 = 1+c^a_4 (F^{\prime 2} -  \frac{s^2}{r^2})} \; , \quad
& \displaystyle{b_0 = 1+c^a_4 \frac{s^2}{r^2}} \; , \nonumber 
\end{eqnarray}
with $s=\sin F$ and $c=\cos F$.
The corresponding equation of motion is
\begin{eqnarray}
\label{soldef2}
h_{ab}^2 \eta_b=\frac{f_\pi}{\Theta^2} && \left\{  \left[ sc
a_L - c_4^a \frac{1}{r^2}
(r^2 F^{\prime} s^2)^{\prime} \right] 
\hat r_a (\mbox{\boldmath$ \hat{r} \times R $})^2 -s a_T
(R_a-\hat r_a (\mbox{\boldmath$\hat{r}R$}))
(\mbox{\boldmath$\hat{r} R$}) \right\} \; ,
\end{eqnarray}
where the differential operator $h_{ab}^2$ is obtained by expanding the
adiabatic lagrangian to second order in the time independent
deformations and where the
angular velocities are replaced by the angular momenta 
$\mbox{\boldmath$R$} = - \Theta \mbox{\boldmath$ \Omega$}^R$
with the moment of inertia $\Theta$. This equation of motion (\ref{soldef2})
is identical to the conservation of the vector current
\be   
\label{vcc}
-\partial^i V_i^a = \partial_i V_i^a  =  \dot V_0^a =
\frac{f_\pi^2}{\Theta^2} s^2 b_T D_{ap}
(\mbox{\boldmath$ \hat{r} \times R$})_p (\mbox{\boldmath$ \hat{r} R$})
\; .   
\ee
In the 
asymptotical region the equation of motion (\ref{soldef2})
may be solved analytically 
\be
\label{defas}
\mbox{\boldmath$\eta$}\stackrel{r \to \infty}{=}
\frac{3g_A}{16 \pi f_{\pi} \Theta^2} e^{-m_{\pi}r}\left[ \mbox{\boldmath$\hat{r}$}
(\mbox{\boldmath$R$})^2-\mbox{\boldmath$R$} 
(\mbox{\boldmath$ \hat{r}R$}) \right] \; ,
\ee
in accordance with ref.\cite{dhm94}. The full
equations induce a monopole and a quadrupole deformation
\begin{eqnarray}
\label{deform}
\eta_L &=& \frac{1}{6f_\pi \Theta^2} \left[
2f(r) \mbox{\boldmath$R$}^2+u(r)
(\mbox{\boldmath$R$}^2-3(\mbox{\boldmath$\hat{r} R$})^2) \right],  \\
\mbox{\boldmath$\eta$}_T &=& -\frac{1}{2f_\pi \Theta^2}
v(r)(\mbox{\boldmath$R$}-\mbox{\boldmath$\hat{r}$} 
(\mbox{\boldmath$ \hat{r}R$}))
(\mbox{\boldmath$ \hat{r}R$}) \; .\nonumber
\end{eqnarray}
The resulting system of differential equations
for the radial functions $f(r)$, $u(r)$ and $v(r)$ is given in the
appendix and
has to be solved numerically subject to the boundary conditions $f(0)=u(0)=v(0)=0$
and
\be   
\label{betafluca}
f(r) \stackrel{r \to \infty} = u(r) \stackrel{r \to \infty} {=} 
v(r) \stackrel{r \to \infty} {=} \frac{3g_A}{8\pi} e^{-m_\pi r} \; , 
\ee
compare (\ref{defas}).
The radial functions are depicted in fig.\ref{radial}. The rotationally
induced soliton deformations (\ref{deform}) are now fixed and enter
into the spatial components of the vector current (see appendix).

\subsection{Transition form factors}

First we evaluate the $M1$
transition operator (\ref{mupo}) by inserting the vector current
with the soliton deformations (\ref{deform}) included
\begin{eqnarray}
\label{m1}
M^{M1}_\lambda (q^2) = -&&\frac{3 \lambda}{2} \biggl\{ 
\frac{2f^2_\pi}{3} \int d^3\!r\, j_1 \frac{s^2}{r} b_T
D_{3 \lambda} \nonumber \\ 
&& \quad +f_\pi \int d^3\!r\, j_1 
\biggl[ \left( \frac{2sc}{r} a_L \eta_L + 
2c_4^a \frac{F^{\prime}s^2}{r} \eta_L^{\prime} 
-s a_0 \mbox{\boldmath$\nabla \eta$}_T \right)
D_{3p} (\delta_{p \lambda} - \hat{r}_p \hat{r}_\lambda) \nonumber \\ 
&& \qquad \qquad \qquad \qquad \qquad -\frac{s}{r} a_L
D_{3p} (\hat{r}_p \eta_{T \lambda} - \eta_{Tp} \hat{r}_\lambda)
\biggr] \biggr\} \\  
= -&&\frac{3 \lambda}{2}   \biggl\{ 
\frac{2f^2_\pi}{3} \int d^3\!r\, j_1 \frac{s^2}{r} b_T 
D_{3 \lambda} \nonumber \\
&& + \frac{1}{45 \Theta^2} \int d^3\!r\, j_1  
\biggl[ \frac{2sc}{r} a_L (10f-u) + 2c_4^a \frac{F^{\prime}s^2}{r}
(10f^{\prime}-u^{\prime}) - \frac{3s}{r} a_0 v \biggr]
\frac{1}{2} \{ D_{3\lambda},\mbox{\boldmath${R}$}^2 \} \nonumber \\  
&& + \frac{1}{15 \Theta^2} \int d^3\!r\, j_1  
\biggl[ \frac{2sc}{r} a_L u  
+ 2c_4^a \frac{F^{\prime}s^2}{r} u^{\prime} + \frac{3s}{r}
a_0 v \biggr]
L_3 R_\lambda \biggl\} \; .  \nonumber
\end{eqnarray} 
Due to the soliton deformations there appear three different operators
$D_{3 \lambda}, \; \frac{1}{2} \{ D_{3\lambda},\mbox{\boldmath${R}$}^2
\}, \;$
$L_3 R_\lambda$ in collective coordinate space.
For that reason the elastic isovector magnetic
form factor and the $M1$ transition form factor are no longer related
by the model independent formula $G^{N \Delta}_{M1} = \sqrt{2} G_M^V$.
Instead we obtain for the elastic isovector magnetic form factor
\begin{eqnarray}
\label{fofav}
\tilde G^V_M(q^2) = && \frac{M_N}{q}  \biggl\{ 
\frac{2f_\pi^2}{3}   \int d^3\!r\, j_1 \frac{s^2}{r} b_T  \\ 
&&  \quad +\frac{1}{30 \Theta^2}
\int d^3\!r\, j_1 \frac{s}{r} [ 
2c a_L (5f+u)
+2c_4^a F^{\prime} s (5f^{\prime}+u^{\prime}) +3 a_0
v ] \biggr\} \; , \nonumber
\end{eqnarray}
and for the $M1$ transition form factor
\begin{eqnarray}
\label{fofam1}
\tilde G^{N \Delta}_{M1}(q^2) = && \sqrt{2} \frac{M_N}{q}  \biggl\{ 
\frac{2f_\pi^2}{3} \int d^3\!r\, j_1 \frac{s^2}{r} b_T  \\
&&  \qquad \; +\frac{1}{20 \Theta^2}
\int d^3\!r\, j_1 \frac{s}{r} [ 2c a_L (10f-u)
+2c_4^a F^{\prime} s (10f^{\prime}-u^{\prime}) -3 a_0
v ] \biggr\} \; . \nonumber
\end{eqnarray}
It is noticed that the factors which multiply the contributions of the
induced components are different.

For the $E2$ transition the soliton
deformations (\ref{deform}) are essential, without them the operator vanishes.
Instead of inserting the rotationally induced components directly into
the expression for the $E2$ multipole operator (\ref{mupo}) we may employ
partial integration and vector current conservation (\ref{vcc})
\begin{eqnarray}
\label{e2}
M^{E2}_\lambda (q^2) = \frac{1}{iq} \sqrt{\frac{5 \pi}{3}}
&& \int d^3\!r\,  \left[
(3j_2-qrj_3) \partial_i V_i^3 - q^2 j_2 x_i V_i^3 \right] Y_{2 \lambda}
\nonumber \\
= \frac{1}{iq} \sqrt{\frac{5 \pi}{3}} && \int d^3\!r\, \biggl\{ 
(3j_2-qrj_3) \dot V_0^3  + f_\pi q^2 j_2 D_{3p} \\ 
&&   \qquad \quad \biggl[ b_0 (
r F^{\prime} c \mbox{\boldmath$\hat{r}$} \times \mbox{\boldmath$\eta$}
- r s \mbox{\boldmath$\hat{r}$} \times \mbox{\boldmath$\eta$}^{\prime})
+c_4^a \frac{F^{\prime} s^2}{r} 
(\mbox{\boldmath$r$} \times \mbox{\boldmath$\nabla$}) \eta_L
\biggr]_p Y_{2 \lambda}  \biggr\} \nonumber \\ 
= \sqrt{\frac{5 \pi}{3}} \frac{q_0}{q \Theta}  && \int d^3\!r\, \biggl\{
f^2_\pi (3j_2-qrj_3) s^2 b_T \nonumber \\
&& -\frac{q^2 j_2}{2} \biggl[ b_0 (
r F^{\prime} c v -rs v^{\prime} ) +2 c_4^a \frac{F^{\prime} s^2}{r} u
\biggr] \biggr\} D_{3p} (R_p - \hat{r}_p (\mbox{\boldmath$\hat{r}R$})
) Y_{2 \lambda} \; .
\nonumber
\end{eqnarray}
In the last step it was noticed that the entire operator may be
written as a total time derivative which in the end may be replaced by
$iq_0$ (compare the derivation of Siegert's theorem \cite{s37}). 
The $E2$ transition form factor may now be computed
according to (\ref{fofa})
\begin{eqnarray}
\label{fofae2}
\tilde G^{N \Delta}_{E2}(q^2) = - \frac{\sqrt{2}}{9q^2 \Theta}  && \left\{ 
f^2_\pi  \int d^3\!r\, (3j_2-qrj_3) s^2 b_T \right. \\ 
&& \left. + \frac{q^2}{2} \int d^3\!r\, j_2 
\left[ b_0 (
r F^{\prime} c v -rs v^{\prime} ) +2 c_4^a \frac{F^{\prime} s^2}{r} u
\right] \right\} \; . \nonumber
\end{eqnarray}
Evidently, this does not exactly coincide with 
the quadrupole form factor because of 
additional contributions from the soliton deformations. 

Finally we evaluate the $C2$ transition operator (\ref{mupo})
\be
\label{c2}
M^{C2}_\lambda (q^2) = -\sqrt{20\pi} \frac{f^2_\pi}{\Theta} 
\int d^3\!r\, j_2 s^2 b_T
D_{3p} (R_p - \hat{r}_p (\mbox{\boldmath$\hat{r}R$})
) Y_{2 \lambda} \; 
\nonumber
\ee
together with the corresponding form factor (\ref{fofa})
\be
\label{fofac2}
\tilde G^{N \Delta}_{C2}(q^2) = -\frac{\sqrt{2}}{3q^2 \Theta}  
f^2_\pi  \int d^3\!r\, j_2 s^2 b_T \;  , 
\ee
which is what naively would be considered the quadrupole form factor. 
It is noticed that the $E2$ and $C2$
form factors coincide in the limit $q_0=q \to 0$ which is indeed Siegert's
theorem \cite{s37}.

\section{Results}

\subsection{Parameters of the model}

For the effective action we use the standard Skyrme model with
$f_\pi = 93MeV$ and $m_\pi=138MeV$. 
\begin{itemize}
\item The Skyrme parameter $e=3.86$ is chosen such that the isovector
magnetic moment fits its experimental value $\mu^V=2.35$ nuclear
magnetons. With this choice eq.(\ref{fofam1}) together with 
(\ref{phomom}) leads to an $M1$ transition form factor at the
photon point $G^{N \Delta}_{M1}(Q^2=0)=3.11$ which meets exactly the
experimental value for the transition amplitude
$M1=\sqrt{\pi \alpha / k_\Delta} q_\Delta / M_N G_{M1}^{N \Delta} (Q^2=0)$
quoted by the particle data group \cite{pdg94} (see table 1). 
\item Nonminimal couplings to vector mesons are incorporated into a
common factor 
$$
\Lambda (q^2) = \lambda \frac{m_V^2}{m_V^2+q^2} + (1-\lambda) \; ,
\eqno{\rm{(4.1)}}
$$
($m_V=770MeV$) to be multiplied with the pure Skyrme model form factors.
The choice $\lambda=0.55$ results in an 
acceptable fit (for this simple model) to the elastic magnetic
proton form factor over the low momentum region.
Of course, it would be preferable to
have the resonances as genuine dynamical fields properly included.
Given a lagrangian which comprises $\pi, \rho, \omega, a1,$
(may be even $\sigma$) mesons and photons in chiral and gauge invariant way, 
this is a straightforward although very tedious task. 
Particularly because it is the point of this paper to emphasize
the importance of rotationally induced $1/N_C$ components which deviate
from the usual hedgehog form of the solitons. It is not an impossible
task. But in view of the fact that such a lagrangian will contain numerous
coupling terms with poorly determined coupling constants we do not
consider this large effort worthwhile because the results will hinge on
the choice of too many parameters. (For elastic proton form factors it
has been
demonstrated that with suitable parameters perfect fits are possible,
see e.g.~\cite{ho96}).
We rather decided to have only one additional parameter which allows to
adjust the e.m. radii when the Skyrme parameter is fixed through the 
magnetic moment. The pure Skyrme model with minimal coupling
cannot fit both observables 
simultaneously (if $f_\pi$ has its exptl. value of 93 MeV).
Although we are confident that the rather poor agreement 
shown in fig.2 for the form factors could be much improved by 
additional flexibility gained in the parameter space of vector meson
models we prefer to demonstrate the essential features with as few parameters 
as possible.
\item The kinematical masses which enter into the prescriptions 
(\ref{relativ}) for the boost to the EV-frame are taken as the 
soliton masses obtained in tree approximation with the rotational
contributions included; the above choice
of parameters yields the values $M_N=1824MeV$ and $M_\Delta=2051MeV$.
\end{itemize}

\subsection{Helicity amplitudes at the photon point}

With the parameters of the model fixed we now are in a position to
calculate transition amplitudes and transition form factors. The values for
the transition moments given by the matrixelements of the
related transition operators at $q^2=0$ turn out to be $\mu^{N
\Delta}=3.73 $ nuclear magnetons and $Q^{N \Delta} = -.037 fm^2$.
The latter value corresponds to a quadrupole moment $-.062 fm^2$ of the
$\Delta^{++}$. 
The above quantities should not be confused with the corresponding 
quantities at the photon point 
$$
G^{N \Delta}_{M1}(Q^2=0)=3.11 \; , \quad
G^{N \Delta}_{E2}(Q^2=0)=-.020 fm^2 \; , \quad
G^{N \Delta}_{C2}(Q^2=0)=-.027 fm^2 \; ,
\eqno{\rm{(4.2)}}
$$
which are considerably smaller. From the form factors at the photon point
helicity amplitudes and electromagnetic ratio may be computed
according to (\ref{helix})-(\ref{ratio}). The transverse amplitudes
and the electromagnetic ratio are compared in table 1 with experimental
data and with a recent calculation in a relativized quark model \cite{wspr90}.

\begin{table}[t]
\begin{center}
\parbox{10.2cm}{\caption{\label{phopodat} 
Photodecay amplitudes and electromagnetic ratio at the photon point
$Q^2=0$.}} 

\begin{tabular}{|c|c|c|c|c|}
& experiment$\quad$ & experiment $\quad$ & relativized quark$\quad$ & Skyrme model$\quad$ \\
& ref. \cite{dmw91}$\quad$ & ref. \cite{pdg94}$\quad$  & model ref.
\cite{wspr90} $\quad$ 
& this work $\quad$\\
\hline
$A_\frac{1}{2} [10^{-3}GeV^{-\frac{1}{2}}]$ & $-135\pm 16\quad$ &
$-141\pm 5\quad$ & $-81\quad$ & $-136\quad$ \\
$A_\frac{3}{2} [10^{-3}GeV^{-\frac{1}{2}}]$ & $-251\pm 33\quad$ &
$-257\pm 8\quad$ & $-170\quad$ & $-259\quad$ \\
\hline
$M1 [10^{-3}GeV^{-\frac{1}{2}}]$ & $~~285\pm 37\quad$ &
$~~293\pm 9\quad$ & $~~188\quad$ & $~~292\quad$ \\
$E2 [10^{-3}GeV^{-\frac{1}{2}}]$ & $-4.6\pm 2.6\quad$ &
$~-3.7\pm 0.9\quad$ & $-8.6\quad$ & $-6.8\quad$ \\
\hline
$E2/M1 [\%]$ & $~-1.57\pm .75\quad$ &
$-1.5\pm .4 \quad$ & $-4.6\quad$ & $-2.3\quad$ \\
\end{tabular}
\end{center} 
\end{table}

Although this comparison shows that the Skyrme model 
reproduces the experimental transverse amplitudes with remarkable
accuracy, it should be mentioned that especially the calculated 
electromagnetic ratio is quite sensitive to parameter changes. 
Therefore the precise number listed in
table I for E2/M1 should perhaps not be taken too seriously, but  
a value of E2/M1 $\simeq -2 \%$ may be accepted as a reliable estimate. 
On the other hand, a comparison of the calculated quantity with
numbers extracted from experimental data is subject to severe
ambiguities because the $\Delta$ resonance is embedded in
the pion-nucleon continuum and this background affects the
exerimental data particularly in the case of the small $E2$ and $C2$
amplitudes. The question of how to extract properties of the isolated
resonance is nontrivial and has lead to various theoretical
investigations \cite{o74} \cite{kmo84} \cite{l88}. One possibility 
may be to compare our calculated model quantity to the value 
$E2/M1=-3.5\%$ obtained by subtracting the background from 
the recent MAMI photo-production data \cite{hdt96}.

With the same reservations we obtain from $G^{N \Delta}_{C2}$ (4.2) 
for the longitudinal amplitude at the photon point
$S_{\frac{1}{2}}=.011 GeV^{-\frac{1}{2}}$ which corresponds to a ratio
$C2/M1=-2.7\%$ comparable in size to the electomagnetic one. This ratio
is depicted in Fig.4 as a function of $Q^2$.

Clearly, a complete calculation of the $\gamma N \to \pi N$ reaction
which includes the coupling of the continuous $\pi N$-background 
with the bound $\Delta$ resonance would be highly desirable for
obtaining complex helicity amplitudes for the $N \Delta$ transition. 
Although it is another conceptual advantage of soliton models (as compared to
quark models) that they naturally contain all the ingredients 
necessary for such a calculation, the efforts required go far beyond
the adiabatic procedures described in ~\cite{es86}\cite{swhh89} because, as
we have stressed here, it is just the non-adiabatic rotational effect
in the soliton profiles and in the interaction terms which are crucial
for these amplitudes.

\subsection{Transition form factors of the $\Delta$ resonance}

In Fig.2 we compare the elastic magnetic proton form factor $G_M^P$ and
the $M1$ transition form factor $G_{M1}^{N \Delta}$ as functions of
$Q^2$ with experimental data. The difference of the two form factors
(apart from a less important isoscalar contribution to the elastic
proton form factor) is essentially due to the soliton deformations
induced by the collective rotation which yield different matrixelements
for nucleon states and for nucleon and $\Delta$ states, respectively
(\ref{fofav},\ref{fofam1}). This difference appears with the correct sign
although its size is somewhat underestimated. It has a simple
geometrical interpretation, namely the spatial distribution of 
densities where $\Delta$-states are
involved extend further out to larger radii because of the centrifugal
forces and consequently these form factors fall off more
rapidly as compared to the corresponding nucleon form factors.
The precise shape of both the elastic and the transition form factor
above $Q^2 \sim 1 GeV^2$ is sensitive to the choice of the kinematical
masses in the boost transformation (\ref{relativ}) with (\ref{phomom})
and to the (very small) values of the nonrelativistic form factors
$\tilde G_{M}(q^2)$ at $q^2 \simeq (M_\Delta+M_N)^2$. Specifics of the
model (e.g. a sixth-order term in the chiral lagrangian, or explicit 
inclusion of dynamical vector mesons) and quantum corrections \cite{mw96}
are known to strongly influence these features. Therefore the
predictive power of a specific model in tree approximation is quite
poor in this respect and can only indicate general features.

In Fig.3 we display the $E2$ (solid line) and $C2$ (dashed line)
form factors. In this plot both form factors are normalized to unity at
the photon point $Q^2=0$. It is noticed that the $Q^2$-dependence of
these two form factors is quite different: while the $C2$-quadrupole
form factor related to the time component of the vector current falls off
smoothly, the $E2$ form factor related to the spatial components changes
sign at $Q\simeq 2.6GeV$. Again details of the shape above $Q^2\sim 1
GeV^2$  depend on the choice of a specific model and on quantum corrections.

Finally in Fig.4 we compare the calculated $C2/M1$ ratio with Bonn,
NINA, and DESY results taken from ref. \cite{s95}. In contrast to the
new Bonn data point at low momentum transfer (on the very left) our
calculation seems to suggest much smaller values in magnitude
comparable to those of the $E2/M1$ ratio. However, again we should 
remember here, that comparison of the model results to the quoted data 
is problematical.

\section{Summary}
In this note we consider the ${\cal O}(N_C^{-1})$ 
corrections induced by the rotation of the classical hedgehog in
isospace. They are crucial for conservation of the vector current
to ${\cal O}(N_C^{-2})$ and therefore allow for an
evaluation of magnetic, electric and longitudinal transition formfactors 
and moments consistent to that order.
The essential features emerge already in the most simple Skyrme model.
However, for reasonable agreement with the elastic proton magnetic
form factor the Skyrme model has to be augmented by a partial 
photon-vectormeson coupling and relativistic recoil corrections.
For these ingredients we have used only very rough approximations;
they could be replaced by more involved techniques.

Apart from pion decay constant $f_\pi=93 MeV$, pion mass $m_\pi=138
MeV$, vector meson mass $m_V=770 MeV$, taken at their physical values,
the model then contains two parameters: the Skyrme constant $e$ and a mixing
parameter $\lambda$ which allows the coupling to the photon field to be
partially mediated through vector mesons.
We use $e$ to fit the isovector magnetic moment of the nucleon to its
experimental value, and $\lambda$ to adjust the elastic
proton magnetic form factor to the standard dipole fit. All
calculations are done in tree approximation and we could argue that
quantum corrections expected for these oservables are absorbed into
the choice of these two parameters. All results about transition moments and
form factors then are free of additional parameters. 

Comparing with very sophisticated and extremely tedious recent
calculations \cite{bdt88} \cite{ck90} \cite{wspr90}
in quark bag and cluster models the essential results for the
nucleon-$\Delta$ transition in this rather simple soliton model are remarkable:
\begin{itemize}
\item 
The M1 transition moment and both transverse amplitudes at photon point
agree with the presently observed values within the experimental
uncertainties. The rather sensitive ratios $E2/M1$ and $C2/M1$
are obtained as $-2.3\%$ and $-2.7\%$, respectively,
for the chosen parameter set.
\item
The M1 transition form factor decreases significantly faster as
function of $Q^2$ than the elastic magnetic form factor.   
\item 
There is a sign change predicted in the E2 transition form factor
around 2-3 $GeV^2$. Its precise location is not very well defined in
this calculation because it sensitively depends on the kinematical mass
in the boost transformation, which is subject to large loop corrections.
The shapes of the E2 and C2 transition form factors are significantly
different from each other.
\item
The $C2/M1$ ratio tentatively follows the experimental data as a 
function of $Q^2$.   
\end{itemize}
The origin of all these results are the rotationally induced monopole
and quadrupole 
deformations of the Skyrme hedgehog. Naturally, these are crucial for
the structure of the $\Delta$ resonance which in the Skyrme soliton model
is an iso-rotational excited state. They are two $N_C$-orders down as
compared to the leading parts of the currents. This is in contrast to
the transition amplitudes for higher nucleon resonances like the
P11(1440), D13(1520), F15(1680), etc. which correspond to time-dependent
small-amplitude fluctuations of the Skyrme hedgehog and therefore are
suppressed only by one $N_C$ order. Their photo-excitation 
amplitudes have been evaluated previously in different versions of the
soliton model without \cite{es86} and with \cite{swhh89} 
inclusion of vector mesons.

\begin{appendix}
\section{}

Here we list the differential equations for the radial functions $f(r),
u(r)$ and $v(r)$ which enter the monopole and quadrupole deformations
\begin{eqnarray}
\eta_L &=& \frac{1}{6f_\pi \Theta^2} \left[
2f(r) \mbox{\boldmath$R$}^2+u(r)
(\mbox{\boldmath$R$}^2-3(\mbox{\boldmath$\hat{r} R$})^2) \right],  
\nonumber \\
\mbox{\boldmath$\eta$}_T &=& -\frac{1}{2f_\pi \Theta^2}
v(r)(\mbox{\boldmath$R$}-\mbox{\boldmath$\hat{r}$} 
(\mbox{\boldmath$ \hat{r}R$}))
(\mbox{\boldmath$ \hat{r}R$}) \; 
\end{eqnarray}
induced by the soliton's rotation. We use the abbreviations (3.4) and
the longitudinal and transverse potentials
\begin{eqnarray}
\displaystyle{
V_L = \frac{2(c^2-s^2)}{r^2} + m^2_\pi c - \frac{2 c^a_4}{r^2}
\biggl[F'^2(c^2-s^2) + \frac{s^2}{r^2} + 2 F''sc - \frac{4 s^2
c^2}{r^2} \biggr]} \nonumber \\
\displaystyle{ V_T = - (F'^2 + \frac{2s^2}{r^2}) + m^2_\pi c -  \frac{c^a_4}{r^2}
\biggl[F''sc  +  F'^2 (2 + s^2)  - \frac{2 s^2 c^2}{r^2} \biggr]} \; .
\end{eqnarray}
The differential equation for the monopole deformation $f(r)$ then becomes
\be
- \frac{1}{r^2} (r^2 b_L f')' + V_L f = 2 f^2_\pi \biggl[sca_L - c^a_4
\frac{1}{r^2} (r^2 F's^2)' \biggr] \; , 
\ee
and similarly those for the quadrupole deformation $u(r)$ and $v(r)$
\begin{eqnarray}
\displaystyle{ - \frac{1}{r^2} (r^2 b_L u')' + \frac{6b_0}{r^2} u 
+ V_L u  -  \frac{6 c
b_L}{r^2} v + \frac{3 c^a_4}{r^2} \biggl( (F'sv)' + 
F'' sv \biggr) } 
&=&  2 f^2_\pi \biggl[ sca_L - c^a_4 \frac{1}{r^2} (r^2 F' s^2)'
\biggr] \nonumber \\  
\displaystyle{ -  \frac{1}{r^2} (r^2 b_0 v')' + \frac{6b_T}{r^2} v + V_T v  -  \frac{4 c
b_L}{r^2} u + \frac{2 c^a_4}{r^2} \biggl( F''su -
F'su' \biggr) }  
&=&  2 f^2_\pi s a_T \; . 
\end{eqnarray}
With these rotationally induced soliton deformations included the
vector current $V^3_i = D_{3p} \tilde V^p_i$ in the intrinsic system
becomes 
\begin{eqnarray}
\tilde V^p_i && =  f^2_\pi \frac{s^2}{r} b_T \varepsilon_{p \ell i} \hat
r_\ell \nonumber \\
&& +  f_\pi b_T \biggl[ s \mbox{\boldmath$\hat{r}$} 
\times \partial_i \mbox{\boldmath$\eta$}_T - F' c
\hat{r}_i \mbox{\boldmath$\hat{r}$} \times
\mbox{\boldmath$\eta$}  + \frac{s(1-c)}{r}
\mbox{\boldmath$\hat{r}$} (\mbox{\boldmath$\hat{r}$}
\times \mbox{\boldmath$\eta$})_i
- sc \partial_i 
\mbox{\boldmath$\hat{r}$} \times (\mbox{\boldmath$\eta$}_T 
+ 2 \mbox{\boldmath$\hat{r}$} \eta_L) \biggr]_p \nonumber \\
&& +  f_\pi c^a_4 (F'^2- \frac{s^2}{r^2}) \biggl[ F' c \hat r_i 
\mbox{\boldmath$\hat{r}$} \times \mbox{\boldmath$\eta$} 
- s \hat{r}_i \mbox{\boldmath$\hat{r}$} \times 
\mbox{\boldmath$\eta$}^{\prime} \biggr]_p 
+  2f_\pi c^a_4 \biggl[ \frac{s}{r} \mbox{\boldmath$\nabla \eta$}_T +
F' \eta^{\prime}_L + \frac{sc}{r^2} \eta_L \biggr] \frac{s^2}{r}
\varepsilon_{ p \ell i} \hat r_\ell \nonumber \\
&& +  f_\pi c^a_4  \frac{s^2}{r^2} \biggl[ (F' c - \frac{s}{r})
\hat{r}_i (\mbox{\boldmath$\hat{r}$} \times
\mbox{\boldmath$\eta$}) - F' \hat{r}_i 
(\mbox{\boldmath$r$} \times \mbox{\boldmath$\nabla$})
\eta_L - \frac{s}{r}
(\mbox{\boldmath$r$} \times \mbox{\boldmath$\nabla$})
\eta_{Ti} - s \mbox{\boldmath$\hat{r}$} \times
\partial_i \mbox{\boldmath$\eta$}_T \biggr]_p \; .
\end{eqnarray}
With the differential equations (A3,A4) it is straightforward to
verify that the vector current $\partial_i V_i^a = \dot V^a_0$ (3.6) is
conserved.

\end{appendix}


\newpage
\begin{figure}
\begin{center} \parbox{8.2cm}{\caption{\label{radial} Radial functions 
$f(r)$ of the monopole deformation and $u(r)$ and $v(r)$ of the 
quadrupole deformation induced by the soliton's rotation
(full lines). For comparison the asymptotical function
is also plotted (dashed
line).}} 
\end{center}
\end{figure}
\begin{figure}
\begin{center} \parbox{8.2cm}{\caption[] {\label{proton} 
Proton magnetic form factor divided by the standard dipole $\mu^P G_D$
and magnetic form factor for the $N \Delta$ transition also divided by the
standard dipole $G_D$ and normalized to one
as obtained from a Skyrme model with $e=3.86$ and vector meson coupling
$\lambda = 0.55$. The data for the proton form factor 
are from the compilations of \cite{h93}
- \cite{a94} , those for the transition form factor are
referenced in \cite{wspr90}. }} 
\end{center}
\end{figure}
\begin{figure}
\begin{center} \parbox{8.2cm}{\caption{\label{electric} 
Electric and scalar form factors for the $N \Delta$ transition divided by the
standard dipole $G_D$ and normalized to one 
as obtained from a Skyrme 
model with $e=3.86$ 
and vector meson coupling $\lambda = 0.55$.
}}
\end{center}
\end{figure}
\begin{figure}
\begin{center} \parbox{8.2cm}{\caption[] {\label{sm}
Ratio $C2/M1$ between the longitudinal and transverse couplings to the 
$\Delta$ resonance. The experimental data are taken from ref. \cite{s95}.
}}
\end{center}
\end{figure}

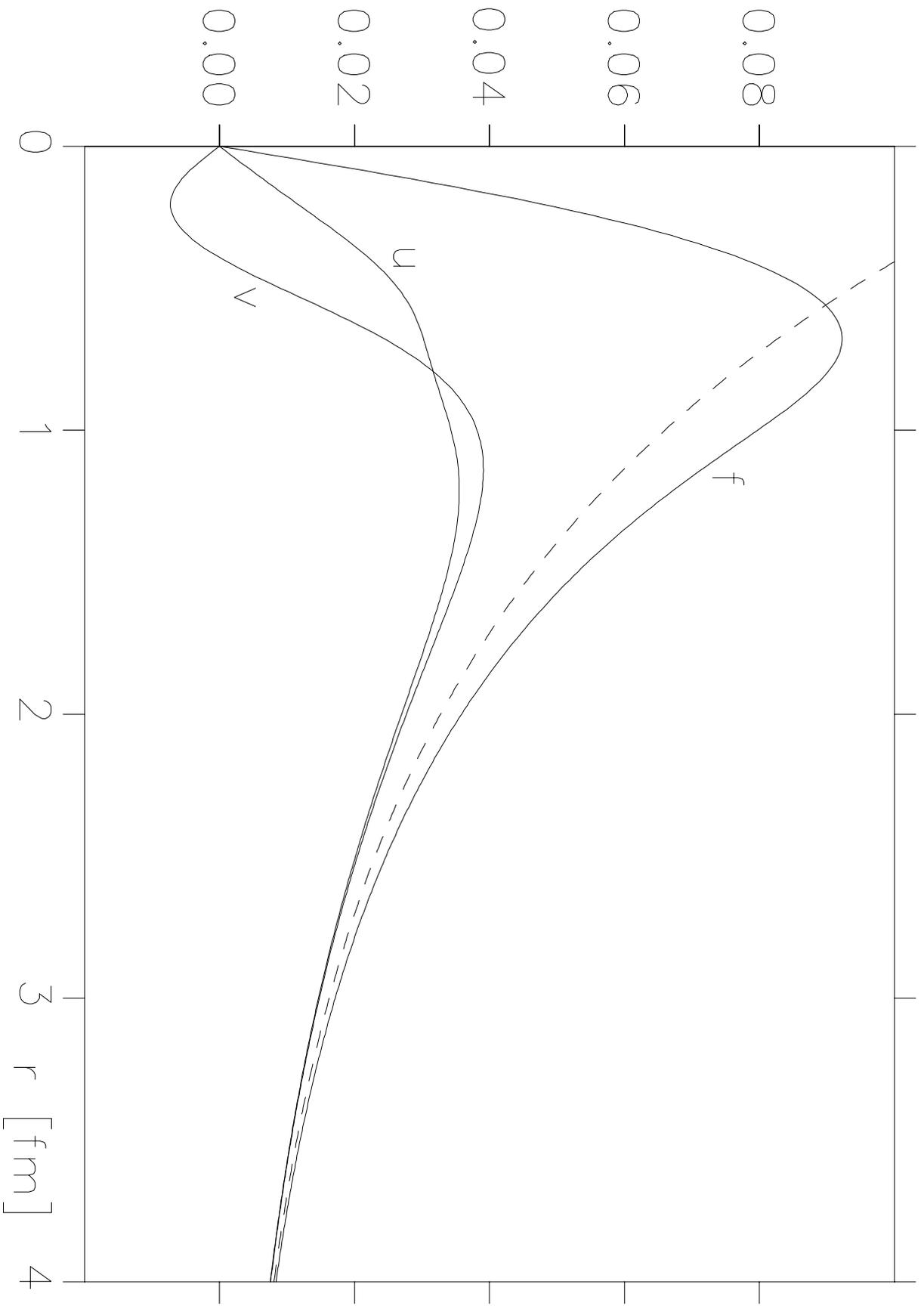
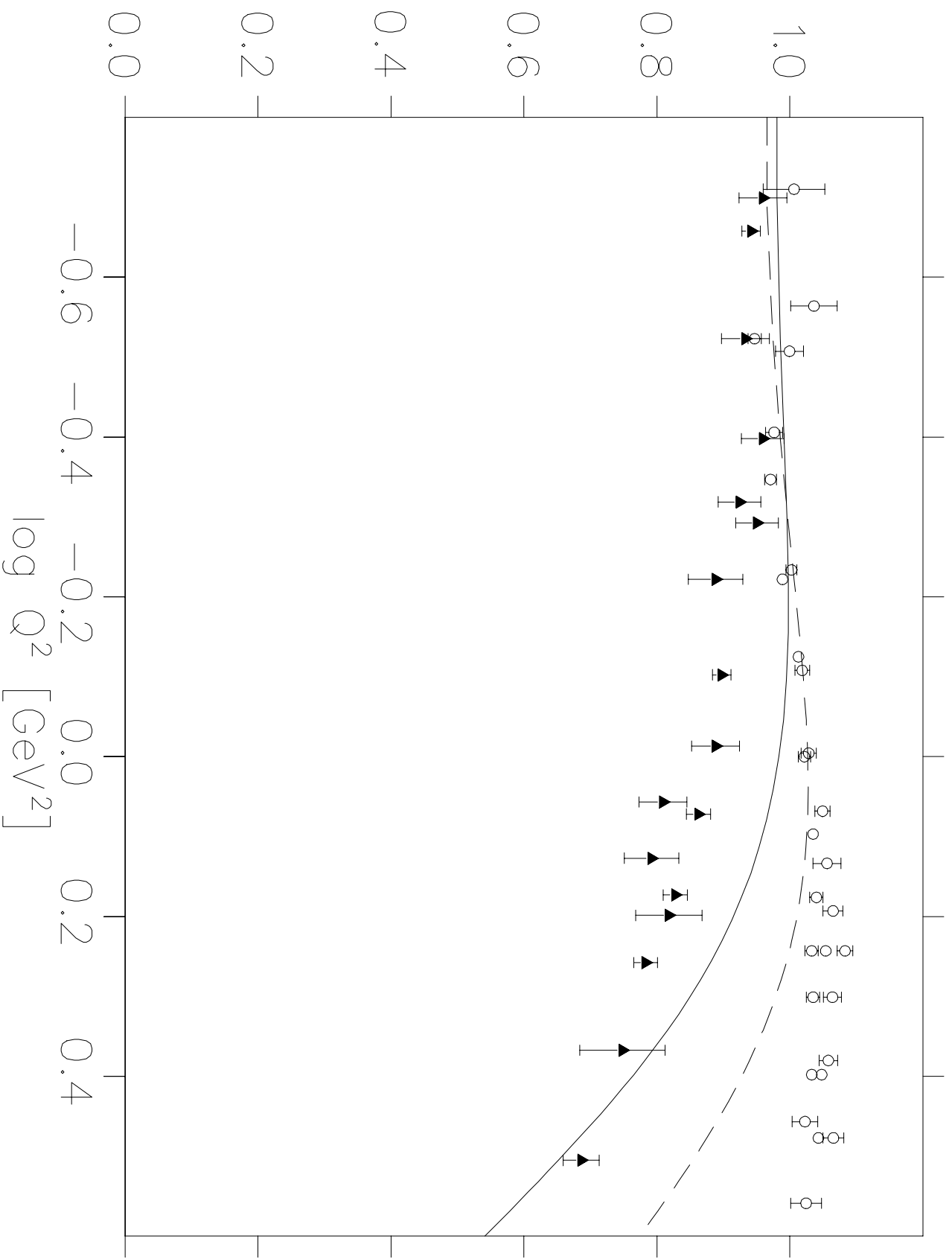
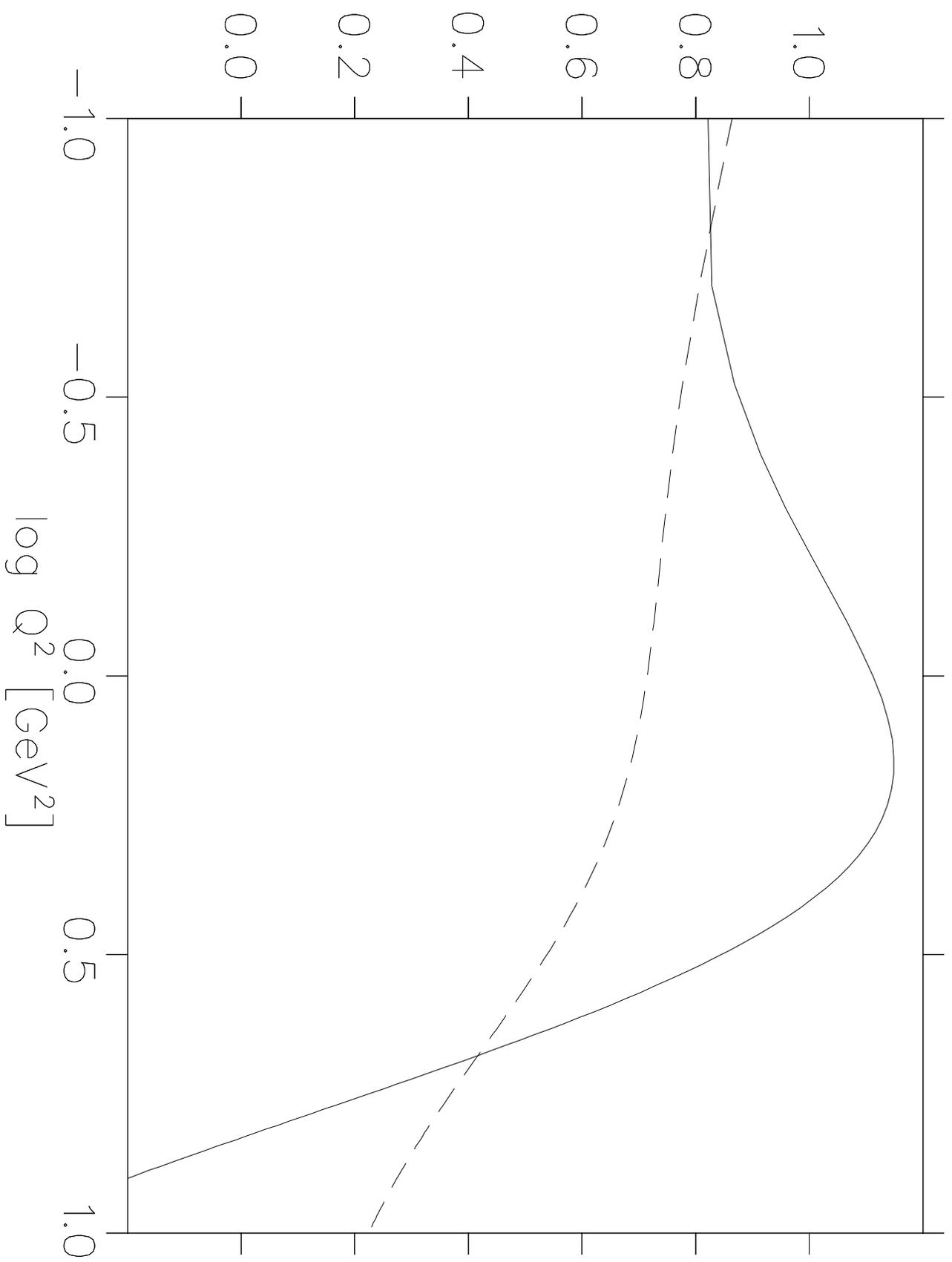
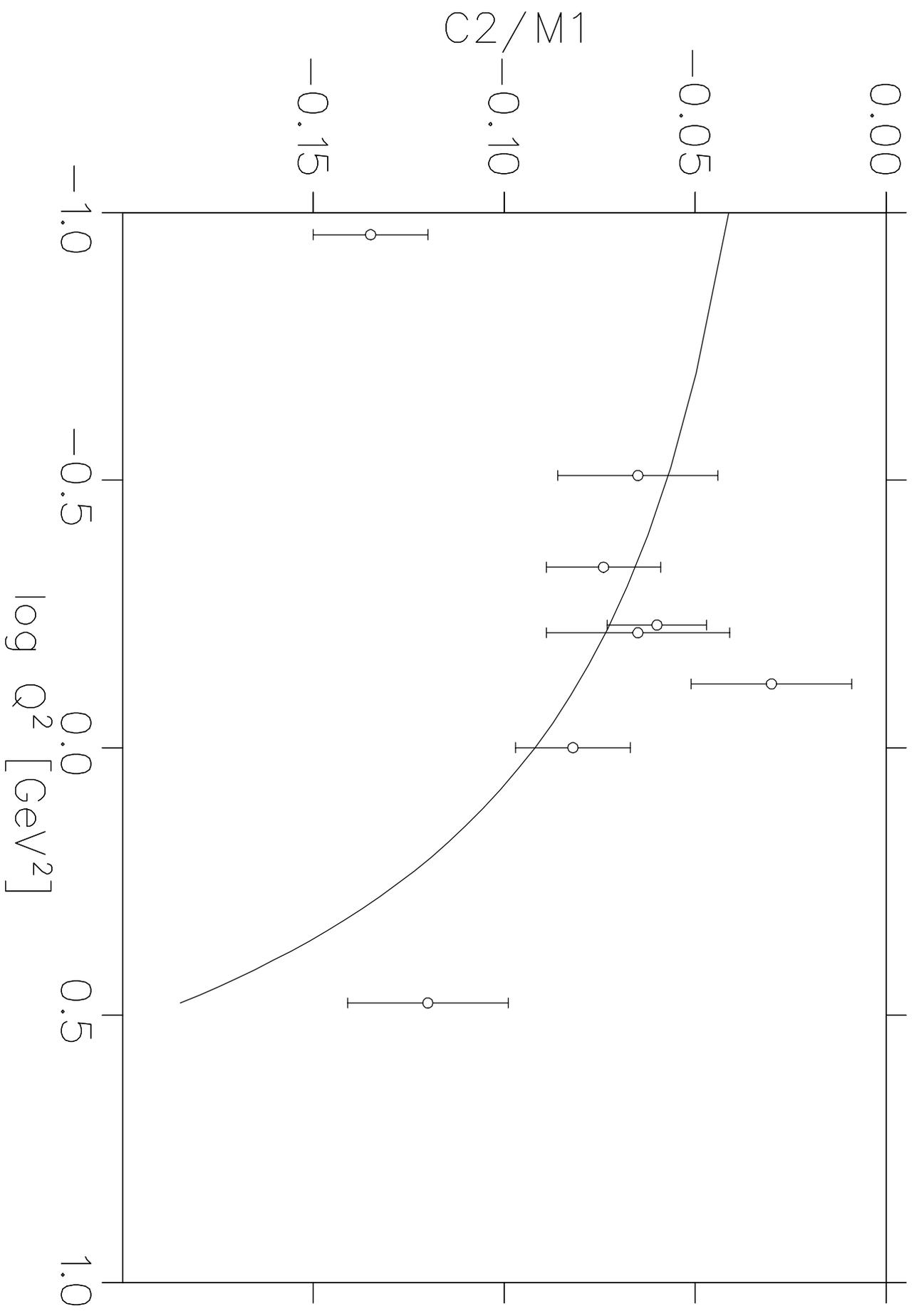
\end{document}